\documentclass[runningheads]{llncs}
\usepackage{graphicx}
\usepackage{tikz}
\usepackage{comment}
\usepackage{amsmath,amssymb} % define this before the line numbering.
\usepackage{color}
\usepackage{hyperref}

\usepackage{multirow}

\usepackage[accsupp]{axessibility}  % Improves PDF readability for those with disabilities

\usepackage{cite}

\begin{document}
% \linenumbers
\pagestyle{headings}
\mainmatter
\def\ECCVSubNumber{8}  % Insert your submission number here

\title{MIPI 2022 Challenge on RGBW Sensor Fusion: Dataset and Report} % Replace with your title

%\begin{comment}
\titlerunning{ECCV-22 submission ID \ECCVSubNumber} 
\authorrunning{ECCV-22 submission ID \ECCVSubNumber} 
\author{Anonymous ECCV submission}
\institute{Paper ID \ECCVSubNumber}
%\end{comment}
%******************

%\begin{comment}
\titlerunning{MIPI 2022 Challenge on RGBW Sensor Fusion}
\author{
Qingyu Yang \and 
Guang Yang \and
Jun Jiang \and
Chongyi Li \and
Ruicheng Feng \and
Shangchen Zhou \and
Wenxiu Sun \and
Qingpeng Zhu \and
Chen Change Loy \and
Jinwei Gu \and
Zhen Wang \and 
Daoyu Li \and
Yuzhe Zhang \and 
Lintao Peng \and 
Xuyang Chang \and 
Yinuo Zhang \and 
Liheng Bian \and
Bing Li \and
Jie Huang \and 
Mingde Yao \and
Ruikang Xu \and
Feng Zhao \and
Xiaohui Liu \and
Rongjian Xu \and 
Zhilu Zhang \and
Xiaohe Wu \and
Ruohao Wang \and
Junyi Li \and
Wangmeng Zuo \and
Zhuang Jia \and
DongJae Lee \and
Ting Jiang \and
Qi Wu \and
Chengzhi Jiang \and 
Mingyan Han \and 
Xinpeng Li \and 
Wenjie Lin \and 
Youwei Li \and
Haoqiang Fan \and 
Shuaicheng Liu
\institute{~}
\vspace{-1cm}
}
\authorrunning{Q. Yang et al.}
\maketitle
\let\thefootnote\relax\footnotetext{\tiny Qingyu Yang$^{1}$ (\email{yangqingyu@sensebrain.site}), Jun Jiang$^{1}$ (\email{jiangjun@sensebrain.site}),  Chongyi Li$^{4}$, Shangchen Zhou$^{4}$, Ruicheng Feng$^{4}$, Wenxiu Sun$^{2,3}$, Qingpeng Zhu$^{2}$, Chen Change Loy$^{4}$, Jinwei Gu$^{1,3}$,   are the MIPI 2022 challenge organizers ($^{1}$SenseBrain, $^{2}$SenseTime Research and Tetras.AI, $^{3}$Shanghai AI Laboratory, $^{4}$Nanyang Technological University). The other authors participated in the challenge. Please refer to Appendix~\ref{appendix:teams} for details.\\ 
\\
MIPI 2022 challenge website: \url{http://mipi-challenge.org/}
}

\begin{abstract}
Developing and integrating advanced image sensors with novel algorithms in camera systems are prevalent with the increasing demand for computational photography and imaging on mobile platforms. However, the lack of high-quality data for research and the rare opportunity for in-depth exchange of views from industry and academia constrain the development of mobile intelligent photography and imaging (MIPI). To bridge the gap, we introduce the first MIPI challenge, including five tracks focusing on novel image sensors and imaging algorithms. In this paper, RGBW Joint Fusion and Denoise, one of the five tracks, working on the fusion of binning-mode RGBW to Bayer, is introduced. The participants were provided with a new dataset including 70 (training) and 15 (validation) scenes of high-quality RGBW and Bayer pairs. In addition, for each scene, RGBW of different noise levels was provided at 24dB and 42dB.
All the data were captured using an RGBW sensor in both outdoor and indoor conditions. The final results are evaluated using objective metrics, including PSNR, SSIM~\cite{ssim}, LPIPS~\cite{lpips}, and KLD. A detailed description of all models developed in this challenge is provided in this paper. More details of this challenge and the link to the dataset can be found at \href{https://github.com/mipi-challenge/MIPI2022}{https://github.com/mipi-challenge/MIPI2022}

\keywords{RGBW, Fusion, Bayer, Denoise, MIPI challenge}
\end{abstract}

\section{Introduction}
\begin{figure}[!ht]
\centering
\includegraphics[width=0.7\textwidth]{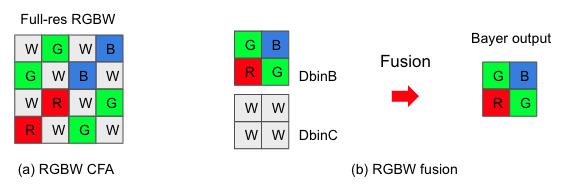}
\caption{The RGBW Fusion task: (a) the RGBW CFA. (b) In the binning mode, DBinB and DBinC are obtained by diagonal averaging of pixels of the same color within a 2$\times$2 window. The joint fusion and denoise algorithm takes DBinB and DBinC as input to get a high-quality Bayer.}
\label{fig:rgbw_cfa}
\setlength{\belowcaptionskip}{0pt plus 3pt minus 2pt}
\end{figure}

% Intro for RGBW sensor
RGBW is a new type of CFA pattern (Fig.~\ref{fig:rgbw_cfa} (a)) designed for image quality enhancement under low light conditions. Thanks to the higher optical transmittance of white pixels over conventional red, green, and blue pixels, the signal-to-noise ratio (SNR) of the sensor output becomes significantly improved, thus boosting the image quality, especially under low light conditions. Recently several phone OEMs, including Transsion, Vivo, and Oppo have adopted RGBW sensors in their flagship smartphones to improve the camera image quality~\cite{oppoRGBW, vivoRGBW, TranssionRGBW}.

The binning mode of RGBW is mainly used in the camera preview mode and video mode, in which the pixels of the same color are averaged in the diagonal direction within a $2\times2$ window in RGBW to further improve the image quality and to reduce the noise. A fusion algorithm is thereby needed to take the input of a diagonal-binning-bayer (DBinB) and a diagonal-binning-white (DBinC) to obtain a Bayer of better signal-to-noise ratio (SNR) in Fig.~\ref{fig:rgbw_cfa} (b). A good fusion algorithm should be able (1) to get a Bayer output from RGBW with least artifacts, and (2) to fully take advantage of the SNR and resolution benefit of white pixels.

The RGBW fusion problem becomes more challenging when the input DBinB and DBinC become noisy especially under low light conditions. A joint fusion and denoise task is thus in demand for real-world applications.

\begin{figure}[!ht]
\centering
\includegraphics[width=\textwidth]{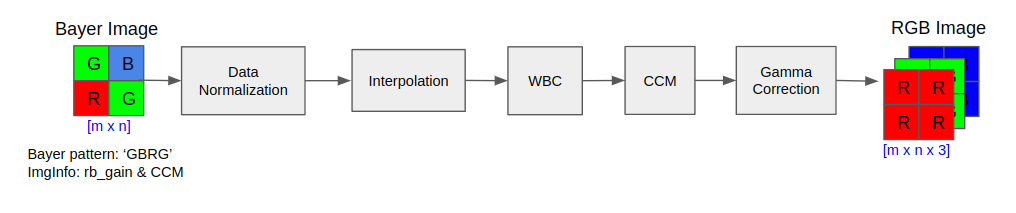}
\caption{An ISP to visuailize the output Bayer and to calculate the loss function.}
\label{fig:simple_isp}
\setlength{\belowcaptionskip}{0pt plus 3pt minus 2pt}
\end{figure}

In this challenge, we intend to fuse the RGBW inputs (DBinB and DBinC in Fig.~\ref{fig:rgbw_cfa} (b)) to denoise and improve the Bayer. The solution is not necessarily deep-learning. However, to facilitate the deep learning training, we provide a dataset of high-quality binning-mode RGBW (DBinB and DBinC) and the output Bayer pairs, including 100 scenes (70 scenes for training, 15 for validation, and 15 for testing). We provide a Data Loader to read these files and show a simple ISP in Fig.~\ref{fig:simple_isp} to visualize the RGB output from the Bayer and calculate loss functions. The participants are also allowed to use other public-domain datasets. The algorithm performance is evaluated and ranked using objective metrics: Peak Signal-to-Noise Ratio (PSNR), Structural Similarity Index (SSIM)~\cite{ssim}, Learned Perceptual Image Patch Similarity (LPIPS)~\cite{lpips}, and KL-divergence (KLD). The objective metrics of a baseline method are available as well to provide a benchmark. 

This challenge is a part of the Mobile Intelligent Photography and Imaging (MIPI) 2022 workshop and challenges emphasizing the integration of novel image sensors and imaging algorithms, which is held in conjunction with ECCV 2022. It consists of five competition tracks:
\begin{enumerate}
  \item RGB+ToF Depth Completion uses sparse and noisy ToF depth measurements with RGB images to obtain a complete depth map.
  \item Quad-Bayer Re-mosaic converts Quad-Bayer RAW data into Bayer format so that it can be processed by standard ISPs.
  \item RGBW Sensor Re-mosaic converts RGBW RAW data into Bayer format so that it can be processed by standard ISPs.
  \item RGBW Sensor Fusion fuses Bayer data and a monochrome channel data into Bayer format to increase SNR and spatial resolution.
  \item Under-display Camera Image Restoration improves the visual quality of the image captured by a new imaging system equipped with an under-display camera.
\end{enumerate}

\section{Challenge}
To develop high-quality RGBW fusion solution, we provide the following resources for participants:
\begin{itemize}
    \item A high-quality RGBW (DBinB and DBinC in Fig.~\ref{fig:rgbw_cfa}b) and Bayer dataset; As far as we know, this is the first and only dataset consisting of aligned RGBW and Bayer pairs, relieving the pain of data collection to develop learning-based fusion algorithms;
    \item A data processing code with Data Loader to help participants get familiar with the provided dataset;
    \item A simple ISP including basic ISP blocks to visualize the algorithm output and to calculate the loss function on RGB results;
    \item A set of objective image quality metrics to measure the performance of a developed solution.
\end{itemize}

\subsection{Problem Definition}
The RGBW fusion task aims to fuse the DBinB and DBinC of RGBW (Fig.~\ref{fig:rgbw_cfa} (b)) to improve the image quality of the Bayer output. By incorporating the white pixels (DBinC) of higher spatial resolution and higher SNR, the output Bayer potentially would have better image quality. In addition, the binning mode of RGBW is mainly used for the preview and video modes in smartphones, thus requiring the fusion algorithms to be lightweight and power-efficient. While we do not rank solutions based on the running time or memory footprint, the computational cost is one of the most important criteria in real applications.

\subsection{Dataset: Tetras-RGBW-Fusion}

\begin{figure}[!ht]
\centering
\includegraphics[width=0.8\textwidth]{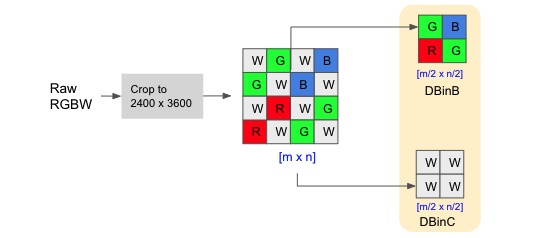}
\caption{Data generation of the RGBW fusion task. The RGBW raw data is captured using an RGBW sensor and cropped to be a size of $2400\times3600$. A Bayer (DBinB) and white (DBinC) image are obtained by averaging the same color in the diagonal direction within a $2\times2$ block.}
\label{fig:data_gen}
\setlength{\belowcaptionskip}{0pt plus 3pt minus 2pt}
\end{figure}

The training data contains 70 scenes of aligned RGBW (DBinB and DBinC input) and Bayer (ground-truth) pairs. For each scene, DBinB at 0dB is used as the ground truth. Noise is synthesized on the 0dB DBinB and DBinC data to provide the noisy input at 24dB and 42dB respectively. The synthesized noise consists of read noise and shot noise, and the noise models are measured on an RGBW sensor. The data generation steps are shown in Fig.~\ref{fig:data_gen}. The testing data contains DBinB and DBinC inputs of 15 scenes at 24dB and 42dB, but the ground truth Bayer results are not available to participants. 
%\vspace{-0.5cm}

\subsection{Challenge Phases}
The challenge consisted of the following phases:
\begin{enumerate}
    \item Development: The registered participants get access to the data and baseline code, and are able to train the models and evaluate their running time locally.
    \item Validation: The participants can upload their models to the remote server to check the fidelity scores on the validation dataset, and to compare their results on the validation leaderboard.
    \item Testing: The participants submit their final results, code, models, and factsheets.
\end{enumerate}

\subsection{Scoring System}
\subsubsection{Objective Evaluation}
The evaluation consists of (1) the comparison of the fused output (Bayer) with the reference ground truth Bayer, and (2) the comparison of RGB from the predicted and ground truth Bayer using a simple ISP (the code of the simple ISP is provided). We use
\begin{enumerate}
    \item Peak Signal-to-Noise Ratio (PSNR)
    \item Structural Similarity Index Measure (SSIM)~\cite{ssim}
    \item Kullback–Leibler Divergence (KLD)
    \item Learned Perceptual Image Patch Similarity (LPIPS)~\cite{lpips}
\end{enumerate}
to evaluate the fusion performance. The PSNR, SSIM, and LPIPS will be applied to the RGB from the Bayer using the provided simple ISP code, while KLD is evaluated on the predicted Bayer directly.

A metric weighting PSNR, SSIM, KLD, and LPIPS is used to give the final ranking of each method, and we will report each metric separately as well. The code to calculate the metrics is provided. The weighted metric is shown below. The M4 score is between 0 and 100, and a higher score indicates a better overall image quality.

\begin{equation}
    \text{M4} = PSNR \cdot \text{SSIM}  \cdot 2^{1-\text{LPIPS}-\text{KLD}} .
\label{eq:M4}
\end{equation}
For each dataset we report the average results over all the processed images belonging to it.

\section{Challenge Results}

Six teams submitted their results in the final phase, and their results have been verified using their submitted code as well. Table.~\ref{tab:results} summarizes the results in the final test phase. \textbf{LLCKP}, \textbf{MegNR}, and \textbf{jzsherlock} are the top three teams ranked by M4 are presented in Eq.~\eqref{eq:M4}, and \textbf{LLCKP} shows the best overall performance. The proposed methods are described in Section \ref{sec:methods}, and the team members and affiliations are listed in Appendix \ref{appendix:teams}.

\begin{table}[!ht]  
    \centering
    
    \begin{tabular}{l | llll | l}
    \hline
        \textbf{Team name} & \textbf{PSNR} & \textbf{SSIM} & \textbf{LPIPS} & \textbf{KLD} &  \textbf{M4}\\ \hline  \hline
        \text{BITSpectral}                                & 36.53         & 0.958             & 0.126      & 0.027         & 63.27 \\ \hline
        \text{BIVLab}         & 35.09         & 0.94              & 0.174      & 0.0255         & 57.98 \\ \hline
        \text{HIT-IIL}                                & 36.66         & 0.958             & 0.128      & 0.02196         & 63.62 \\ \hline
        \text{jzsherlock}                                & 37.05         & 0.958             & 0.132      & 0.29         & \textbf{63.84} \\ \hline
        
        \text{LLCKP}                                              & 36.89         & 0.952               & 0.054      & 0.017           & \textbf{67.07} \\ \hline
        \text{MegNR}         & 36.98         & 0.96              & 0.098      & 0.0156         & \textbf{65.55} \\ \hline
        
    \end{tabular}
    \caption{MIPI 2022 Joint RGBW Fusion and Denoise challenge results and final rankings. PSNR, SSIM, LPIPS, and KLD are calculated between the submitted results from each team and the ground truth data. A weighted metric, M4, by Eq.~\eqref{eq:M4} is used to rank the algorithm performance, and the top three teams with the highest M4 are highlighted.  
    \label{tab:results}}
\end{table}

\begin{figure}[!ht]
\setlength{\abovecaptionskip}{0.cm}
\setlength{\belowcaptionskip}{-0.cm}
\centering
\includegraphics[width=\textwidth]{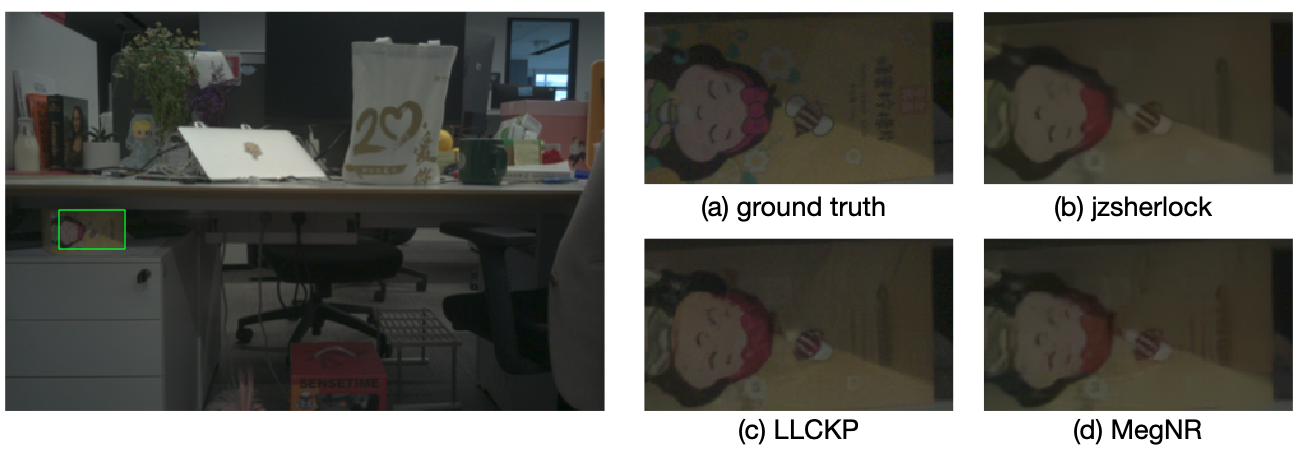}
\caption{Qualitative image quality (IQ) comparison. The results of one of the test scenes (42dB) are shown. While the top three fusion methods achieve high objective IQ metrics in Table.~\ref{tab:results}, details and texture loss are noticeable on the yellow box. The texts on the box are barely interpretable in (b), (c), and (d). The RGB images are obtained by using the ISP in Fig.~\ref{fig:simple_isp}, and its code is provided to participants.}
\label{fig:IQ1}
\end{figure}

\begin{figure}[!ht]
\setlength{\abovecaptionskip}{0.cm}
\setlength{\belowcaptionskip}{-0.cm}
\centering
\includegraphics[width=\textwidth]{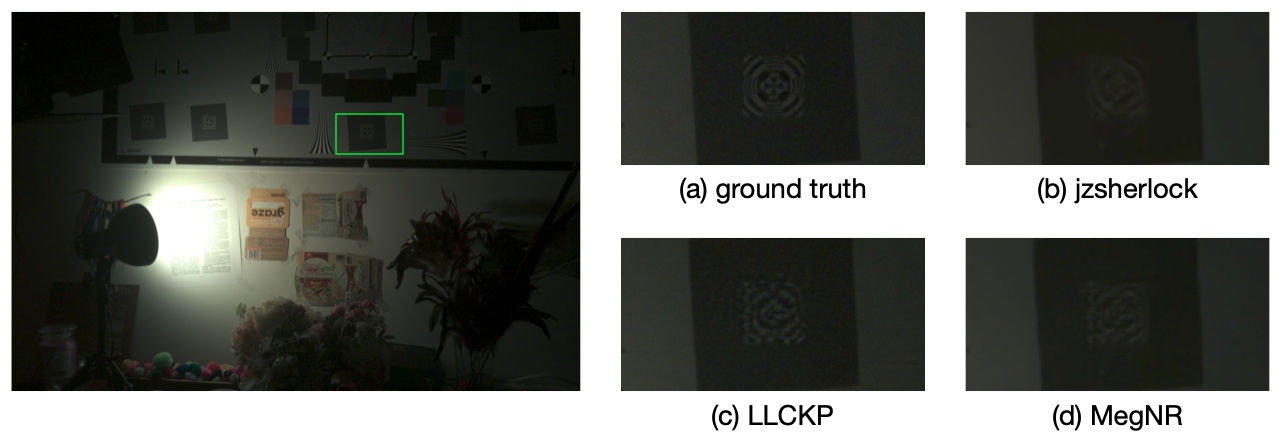}
\caption{Qualitative image quality (IQ) comparison. The results of one of the test scenes (42dB) are shown. Oversmoothing in the top three methods in Table.~\ref{tab:results} can be found when compared with the ground truth. The test chart becomes distorted in (b), (c), and (d). The RGB images are obtained by using the ISP in Fig.~\ref{fig:simple_isp}, and its code is provided to participants.}
\label{fig:IQ2}
\end{figure}

To learn more about the algorithm performance, we evaluated the qualitative image quality in addition to the objective IQ metrics in Fig.~\ref{fig:IQ1} and Fig.~\ref{fig:IQ2} respectively. While all teams in Table.~\ref{tab:results} have achieved high PSNR and SSIM, the detail and texture loss can be found on the yellow box in Fig.~\ref{fig:IQ1} and on the test chart in Fig.~\ref{fig:IQ2}. When the input has a large amount of noise and the scene is under low light conditions, oversmoothing tends to yield higher PSNR at the cost of detail loss perceptually. 

In addition to benchmarking the image quality of fusion algorithms, computational efficiency is evaluated because of the wide adoption of RGBW sensors in smartphones. We measured the running time of the RGBW fusion solutions of the top three teams in Table.~\ref{tab:runtime}. While running time is not employed in the challenge to rank fusion algorithms, computational cost is critical when developing algorithms for smartphones. jzsherlock achieved the shortest running time among the top three solutions on a workstation GPU (NVIDIA Tesla V100-SXM2-32GB). With sensor resolution of mainstream smartphones reaching 64M or even higher, power-efficient fusion algorithms are highly desirable.

\begin{table}[!ht]  
    \centering
    
    \begin{tabular}{l | l | l}
    \hline
        \textbf{Team name} & \textbf{1200$\times$1800 (measured)}  &  \textbf{16M} (estimated)\\ \hline  \hline
        \text{jzsherlock}          & \textbf{3.7s}    &  \textbf{27.4s} \\ \hline
        \text{LLCKP}                & 7.1s                  &  52.6s  \\ \hline
        \text{MegNR}              &  12.4s               &  91.9s \\ \hline

    \end{tabular}
    \caption{Running time of the top three solutions ranked by Eq.~\eqref{eq:M4} in the 2022 Joint RGBW Fusion and Denoise challenge. The running time of input of $1200\times1800$ was measured, while the running time of a 64M RGBW sensor was based on estimation (the binning-mode resolution of a 64M RGBW sensor is 16M).  The measurement was taken on an NVIDIA Tesla V100-SXM2-32GB GPU.
    \label{tab:runtime}}
\end{table}

\section{Challenge Methods}
\label{sec:methods}
In this section, we describe the solutions submitted by all teams paticipanting in the final stage of MIPI 2022 RGBW Joint Fusion and Denoise Challenge. 

\subsection{BITSpectral}
\begin{figure}[!ht]
\setlength{\abovecaptionskip}{0.cm}
\setlength{\belowcaptionskip}{-0.cm}
\centering
\includegraphics[width=0.99\textwidth]{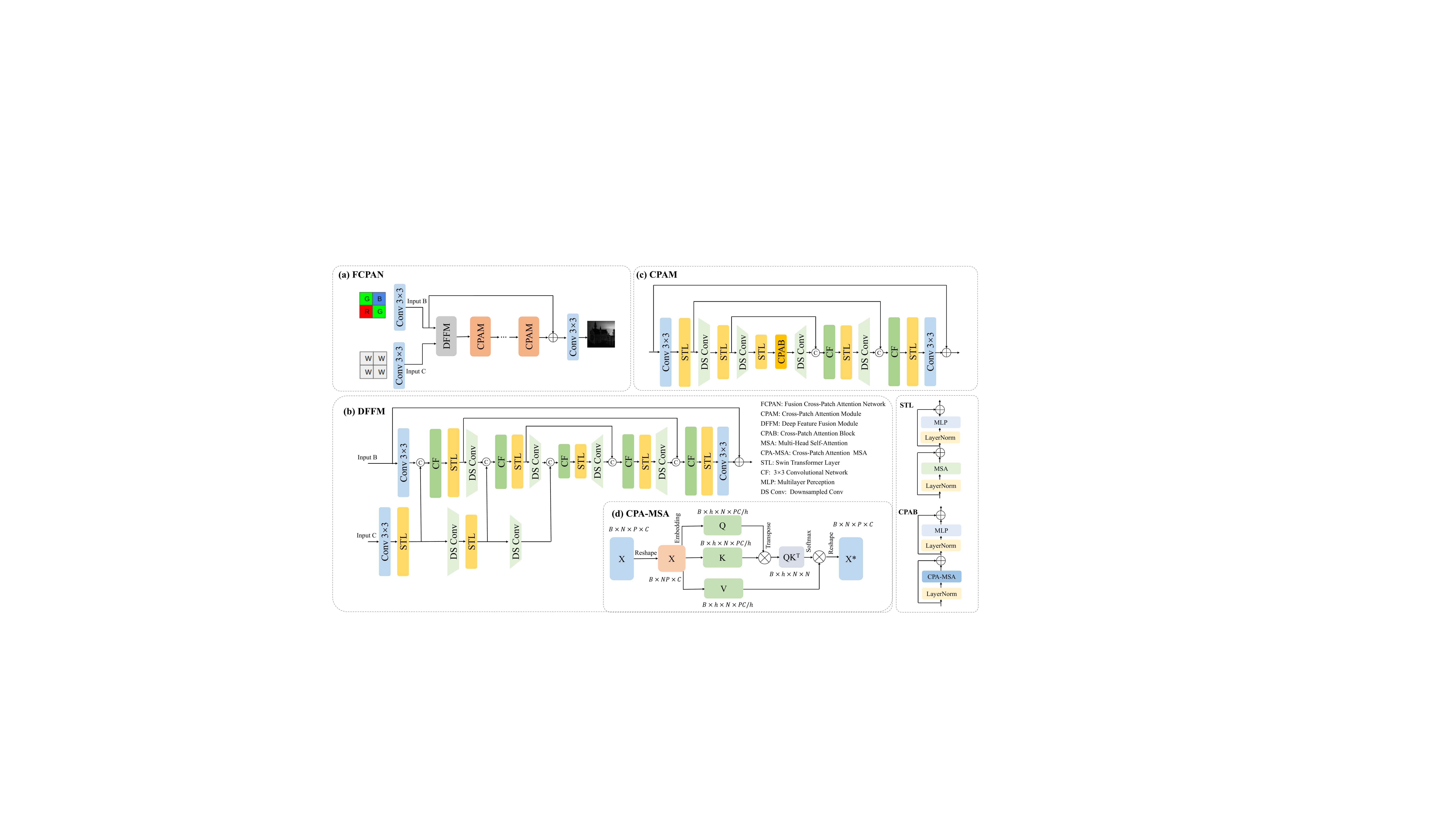}
\caption{The model architecture of BITSpectral.}
\label{fig:nn-BITSpectral}
\end{figure}
BITSpectral developed a transformer-based network, Fusion Cross-Patch Attention Network (FCPAN), for this joint fusing and denoising task. 
The FCPAN is presented in Fig.~\ref{fig:nn-BITSpectral} (a) consisting of a Deep Feature Fusion Module (DFFM) and several Cross-Patch Attention Modules (CPAM).
The input of DFFM contains an RGGB Bayer pattern and a W channel. 
The output of DFFM is the fused features of RGBW, which is fed to CPAM for depth feature extraction.
CPAM is a U-shape network with spatial downsampling to reduce computational complexity.
They proposed to use 4 CPAMs in the network.

Fig.~\ref{fig:nn-BITSpectral} also includes the details of Swin Transformer Layer~\cite{liu2021swin} (STL), the Cross-Patch Attention Block (CPAB), and Cross-Patch Attention Multi-Head Self-Attention (CPA-MSA).
They used STL to extract the attention within feature patches in each stage and CPAB to directly obtain the global attention among patches for the innermost stage.
Compared with STL, CPAB has an extended range of perception due to the cross-patch attention.

\subsection{BIVLab}
\begin{figure}[!ht]
\setlength{\abovecaptionskip}{0.cm}
\setlength{\belowcaptionskip}{-0.cm}
\centering
\includegraphics[width=0.99\textwidth]{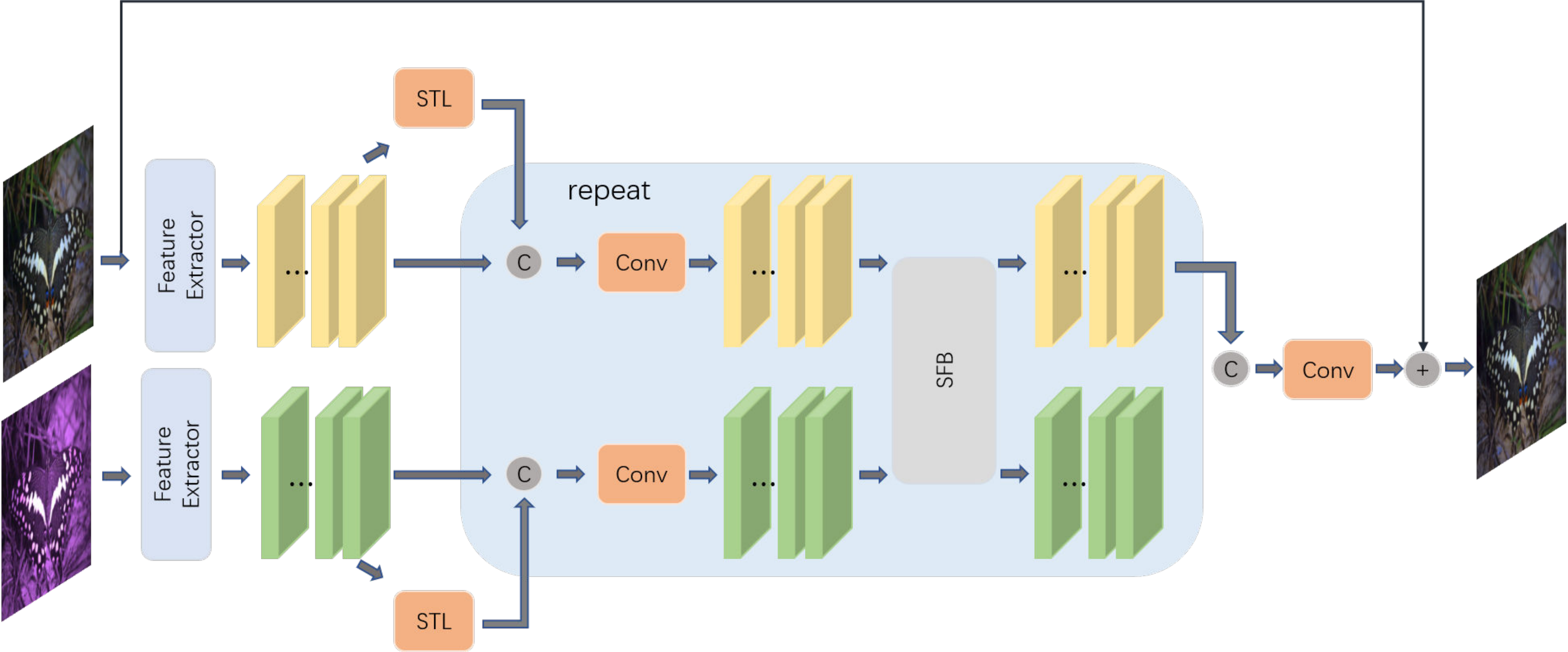}
\caption{The model architecture of BIVLab.}
\label{fig:nn-BIVLab}
\end{figure}
\text{BIVLab} proposed a Self-Guided Spatial-Frequency Complement Network(SG-SFCN) for the RGBW joint fusion and denoise task.
%to reconstruct clean bayer from degraded RGBW DBinB and DBinC bayers. 
As shown in Fig.~\ref{fig:nn-BIVLab}, the swin transformer layer (STL)~\cite{liu2021swin} is adopted to extract rich features from DBinB and DBinC separately.
SpaFre blocks (SFB)~\cite{xu2021deep} then fuses the DBinB and DBinC in complementary spatial and frequency domains.
In order to handle the different noise levels, the features extracted by the STL, which contain the noise-level information, are applied to each SFB as a guidance.
Finally, the denoised Bayer is obtained by adding the predicted Bayer residual to the original DBinB Bayer. 
During the training, all the images are cropped to patches of size $720\times720$ in order to guarantee essential global information.

%the proposed SG-SFCN firstly deploy the swin transformer layer(STL)~\cite{liu2021swin} to extract rich features concerning the noise levels. Then we fuse the DBinB and DBinC bayer in complementary spatial and frequency domains progressively using SpaFre blocks (SFB)~\cite{xu2021deep}, which introduced attention to further facilitate the fusion of information in different bayers. In order to handle the different noise levels, the features extracted by the STL are applied to each SFB as a guidance. Finally, the clean bayer is obtained by adding the predicted bayer residual, that produced by the concatenated features from each SFB, to the original DBinB bayer.

\subsection{HIT-IIL}
\begin{figure}[!ht]
\setlength{\abovecaptionskip}{0.cm}
\setlength{\belowcaptionskip}{-0.cm}
\centering
\includegraphics[width=0.99\textwidth]{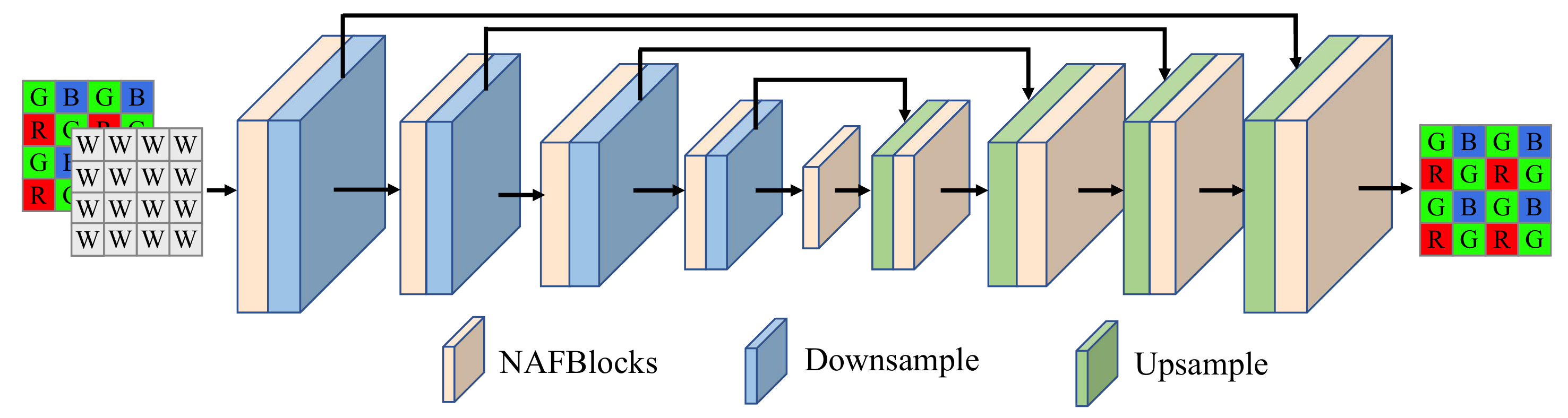}
\caption{The model architecture of HIT-IIL.}
\label{fig:nn-HIT-IIL}
\end{figure}
\text{HIT-IIL} proposed a NAFNet~\cite{chen2022simple} based model for the RGBW Joint Fusion and Denoise task.
As shown in Fig.~\ref{fig:nn-HIT-IIL}, the framework consists of a 4-level encoder-decoder and bottleneck module.
For the encoder, the numbers of NAFNet’s blocks for each level are 2, 2, 4, and 8. 
For the decoder, the numbers of NAFNet’s blocks are set to 2 for all of the 4 levels.
In addition, the bottleneck module contains 24 NAFNet’s blocks.
Unlike the original NAFNet design, the skip connection between the input and the output is removed in their method.

During the training, they also used two data augmentation strategies. 
The first one is mixup, which generates the synthesized images as: 
\begin{equation}
    \centering
    \hat{\mathbf{x}} = \mathbf{a} * \mathbf{x}_{24} + \mathbf{(1-a)} * \mathbf{x}_{42} .
\end{equation}
Here, the $\mathbf{x}_{24}$ and $\mathbf{x}_{42}$ denote the images of the same scene with noise levels of 24dB and 42dB.
A random variable $\mathbf{a}$ is selected between 0 and 1 to generate the synthesised image $\hat{\mathbf{x}}$.
Their second augmentation strategy is the image flip proposed in~\cite{liu2019learning}.

\subsection{jzsherlock}
\begin{figure}[!ht]
\setlength{\abovecaptionskip}{0.cm}
\setlength{\belowcaptionskip}{-0.cm}
\centering
\includegraphics[width=0.99\textwidth]{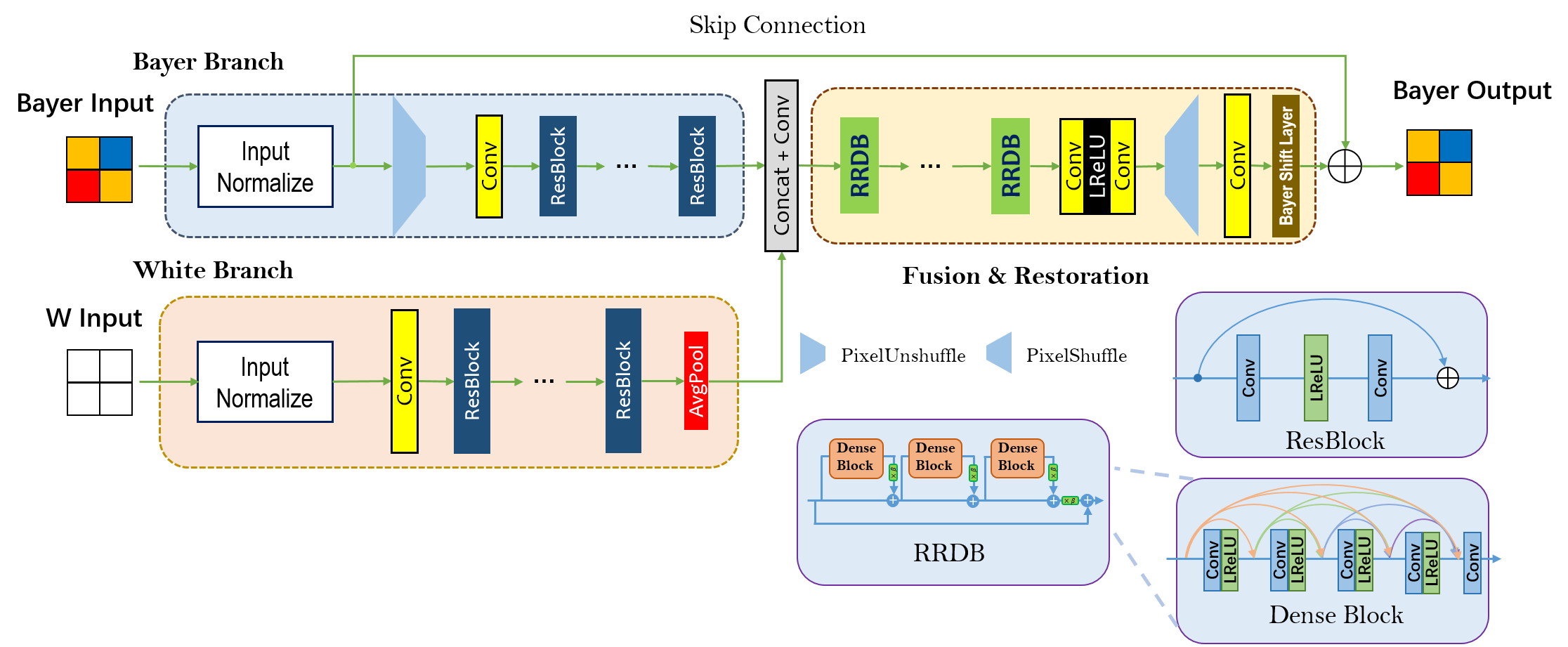}
\caption{The model architecture of jzsherlock.}
\label{fig:nn-jzsherlock}
\end{figure}
Jzsherlock proposed a dual-branch network for the RGBW joint fusion and denoise task. 
The entire architecture, consisting of a Bayer branch and a white branch, is shown in Fig.~\ref{fig:nn-jzsherlock}. 
The Bayer branch's input is a normalized noisy Bayer image and output the denoised result.
After pixel unshuffle operation with scale=2, the Bayer image is converted to GBRG channels.
They use stacked ResBlocks without BatchNorm (BN) layers to extract the feature maps of noisy Bayer image.
On the other hand, the white branch extracts the features from the corresponding white image using stacked ResBlocks as well. 
An average pooling layer rescales the white image features to the same size as Bayer branch for feature fusion.
Several Residual-in-Residual Dense Blocks (RRDB) \cite{wang2018esrgan} are applied to the fused feature maps for restoration. 
After the RRDB blocks, a Conv+LeakyReLU+Conv structure is applied to enlarge the feature map channels by a scale of 4. 
Then pixel shuffle with scale=2 is applied to upscale the feature maps to the input size. 
A Conv layer is used to convert the output to the GBRG 4 channels. 
Finally, a skip connection is applied to add the input Bayer to form the final denoised result. 

The network is trained by L1 loss in the normalized domain. The final normalization with min=64 and max=1023, with values out of the range clipped.

\subsection{LLCKP}
\begin{figure}[!ht]
\setlength{\abovecaptionskip}{0.cm}
\setlength{\belowcaptionskip}{-0.cm}
\centering
\includegraphics[width=0.9\textwidth]{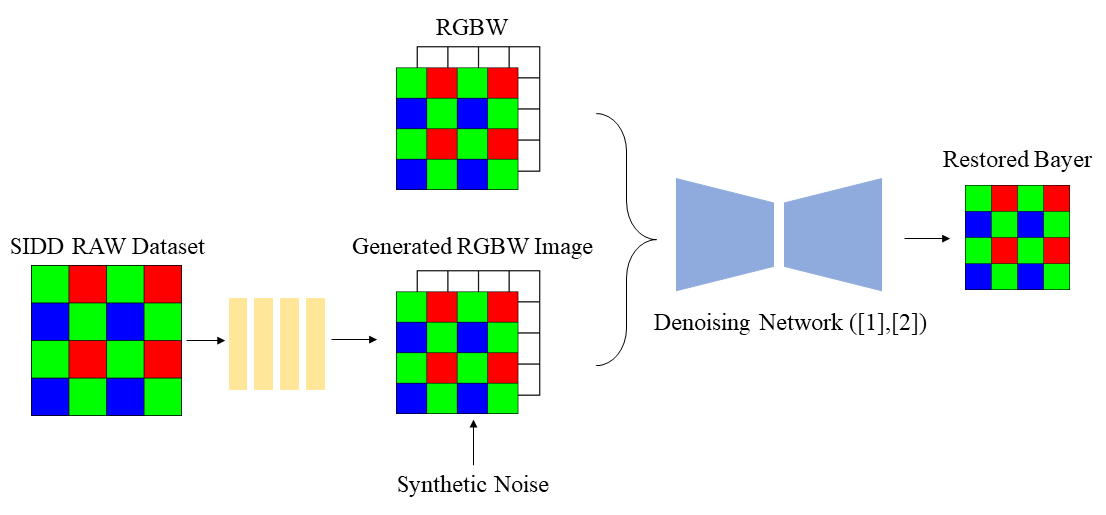}
\caption{The model architecture of LLCKP.}
\label{fig:nn-LLCKP}
\end{figure}
LLCKP proposed a denoising method based on existing image restoration model\cite{zamir2022restormer}, \cite{chen2022simple}. 
As shown in Fig.~\ref{fig:nn-LLCKP}, they synthesized RGBW images from GT GBRG images with additional synthetic noise and real-noise pair (noisy images provided by challenge).
They also used 20,000 pairs of RAW image from SIDD with normal exposure and synthesized RGBW images as extra data.
During the training, the Restormer model's \cite{zamir2022restormer} weights are pre-trained on SIDD RGB images.
Data augmentation~\cite{liu2019learning} and cutmix~\cite{yun2019cutmix} are applied during the training phase.

\subsection{MegNR}
\begin{figure}[!ht]
\setlength{\abovecaptionskip}{0.cm}
\setlength{\belowcaptionskip}{-0.cm}
\centering
\includegraphics[width=0.99\textwidth]{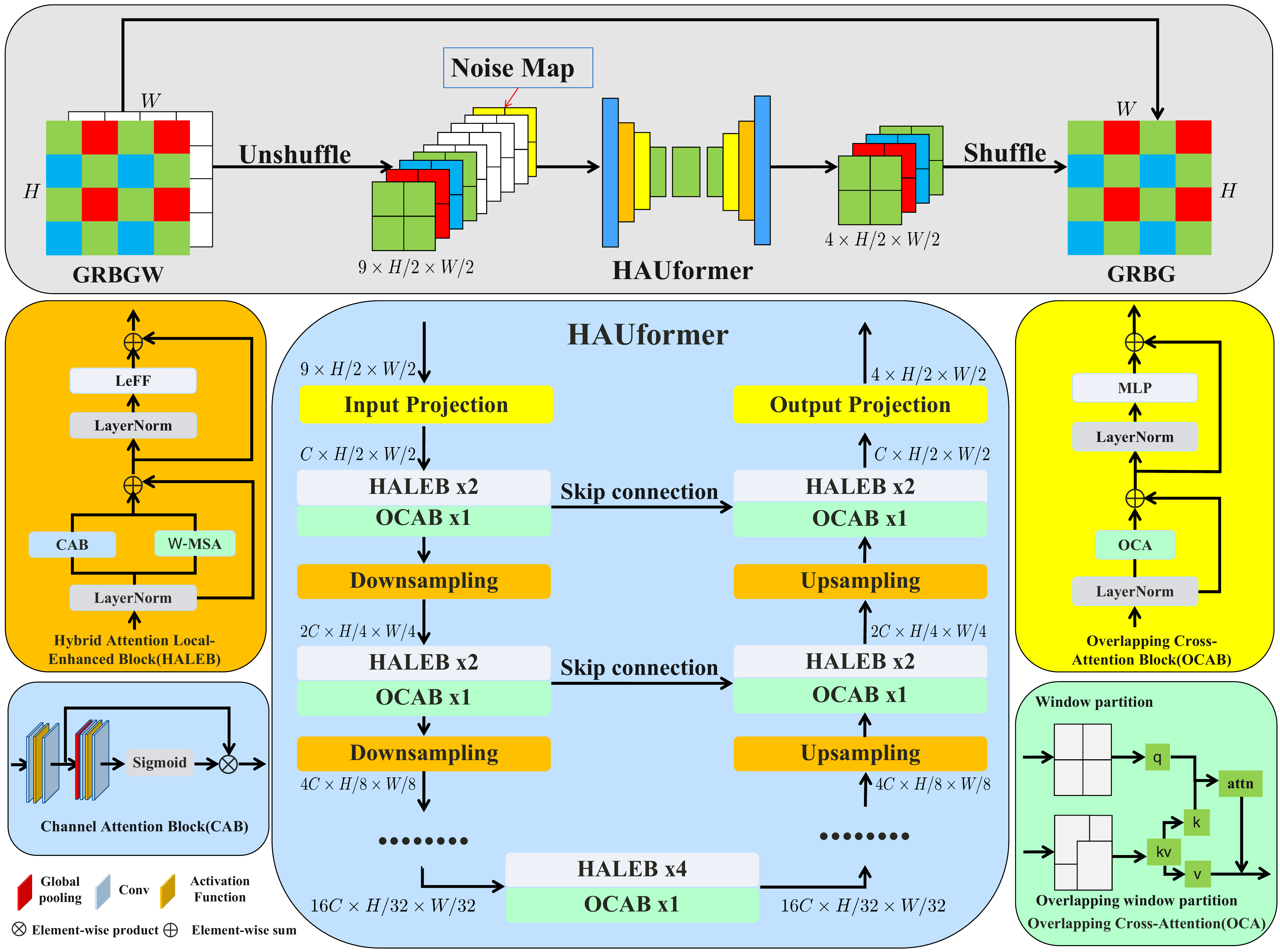}
\caption{The model architecture of MegNR.}
\label{fig:nn-megnr}
\end{figure}
MegNR proposed a pipeline for the RGBW Joint Fusion and Denoise task.
The overall diagram is shown in Fig.~\ref{fig:nn-megnr}.
The pixel-unshuffle(PU)~\cite{sun2022hybrid} is firstly applied to RGBW images to split them into independent channels.
Inspired by Uformer~\cite{wang2022uformer}, they developed their RGBW fusion and reconstruction network, HAUformer.
They replaced the LeWin Blocks~\cite{wang2022uformer} in Uformer's original design and included two modules Hybrid Attention Local-Enhanced Block(HALEB) and Overlapping Cross-Attention Block(OCAB) to capture more long-range dependencies information and useful local context.
Finally, the pixel-shuffle(PS)~\cite{shi2016real} module restored the output to the standard Bayer format.

\section{Conclusions}
In this paper, we summarized the Joint RGBW Fusion and Denoise challenge in the first Mobile Intelligent Photography and Imaging workshop (MIPI 2022) held in conjunction with ECCV 2022. The participants were provided with a high-quality training/testing dataset for RGBW fusion and denoise, which is now available for researchers to download for future research. We are excited to see so many submissions within such a short period, and we look forward for more research in this area.

\section{Acknowledgements}
We thank Shanghai Artificial Intelligence Laboratory, Sony, and Nanyang Technological University to sponsor this MIPI 2022 challenge. We thank all the organizers and participants for their great work. 

\appendix
\section{Teams and Affiliations}
\label{appendix:teams}
\tiny
\textbf{BITSpectral} \\
\textbf{Title}: Fusion Cross-Patch Attention Network for RGBW Joint Fusion and Denoise \\
\textbf{Members}:
Zhen Wang (wzhstruggle@163.com), Daoyu Li, Yuzhe Zhang, Lintao Peng, Xuyang Chang, Yinuo Zhang, Liheng Bian \\
\textbf{Affiliations}:
Beijing Institute of Technology\\
\\
\\
\textbf{BIVLab} \\
\textbf{Title}: Self-Guided Spatial-Frequency Complement Network for RGBW Joint Fusion and Denoise\\
\textbf{Members}:
Bing Li (frigid@mail.ustc.edu.cn), Jie Huang, Mingde Yao, Ruikang Xu, Feng Zhao\\
\textbf{Affiliations}:
University of Science and Technology of China\\
\\
\\
\textbf{HIT-IIL} \\
\textbf{Title}: NAFNet for RGBW Image Fusion\\
\textbf{Members}:
Xiaohui Liu (xh720199@gmail.com), 
Xiaohui Liu, Rongjian Xu, Zhilu Zhang, Xiaohe Wu, Ruohao Wang, Junyi Li, Wangmeng Zuo \\
\textbf{Affiliations}:
Harbin Institute of Technology\\
\\
\\
\textbf{jzsherlock} \\
\textbf{Title}: Dual Branch Network for Bayer Image Denoising Using White Pixel Guidance \\
\textbf{Members}:
Zhuang Jia (jiazhuang@xiaomi.com)\\
\textbf{Affiliations}:
Xiaomi\\
\\
\\
\textbf{LLCKP} \\
\textbf{Title}: Synthetic RGBW image and noise \\
\textbf{Members}:
DongJae Lee (jhtwosun@kaist.ac.kr) \\
\textbf{Affiliations}:
KAIST\\
\\
\\
\textbf{MegNR} \\
\textbf{Title}: HAUformer: Hybrid Attention-guided U-shaped Transformer for RGBW Fusion Image Restoration \\
\textbf{Members}:
Ting Jiang (jiangting@megvii.com), Qi Wu, Chengzhi Jiang, Mingyan Han, Xinpeng Li, Wenjie Lin, Youwei Li, Haoqiang
Fan, Shuaicheng Liu\\
\textbf{Affiliations}:
Megvii Technology\\
\\
\\

\clearpage
\bibliographystyle{splncs04}
\bibliography{egbib}
\end{document}